\documentclass[lettersize,journal]{IEEEtran} % journal vs draftcls, and onecolumn
\IEEEoverridecommandlockouts
\usepackage{amsmath,amssymb,amsfonts}
\usepackage{tabu}
\usepackage{graphicx}
\usepackage{xcolor}
\usepackage{subcaption}
\usepackage{longtable}
\usepackage{booktabs}

\begin{document}

\title{Channel Estimation for Reconfigurable Intelligent Surface-Assisted Full-Duplex MIMO with Hardware Impairments}

\author{\IEEEauthorblockN{Alexander James Fernandes and Ioannis Psaromiligkos} \\
\thanks{{Department of Electrical and Computer Engineering}, 
{McGill University}, Montreal, QC, Canada. Email: 
alexander.fernandes@mail.mcgill.ca; ioannis.psaromiligkos@mcgill.ca}
\thanks{This work was supported in part by the Natural Science and Engineering Research Council of Canada under the Discovery Grant Program and in part by the Vadasz Scholar McGill Engineering Doctoral Award.}
\thanks{Accepted for publication in \emph{IEEE Wireless Communications Letters}.}
\thanks{\textcopyright~2023 IEEE. DOI: 10.1109/LWC.2023.3288517}
}

% The paper headers
\markboth{}%
%\markboth{Name of Journal,~Vol.~\#, No.~\#, Month~Year}%
{}

\maketitle

\begin{abstract}
We consider the problem of channel estimation in a multiple-input-multiple-output (MIMO) full-duplex (FD) wireless communication system assisted by a reconfigurable intelligent surface (RIS) with hardware impairments (HI) occurring at the transceivers and RIS elements.
We propose an unbiased channel estimator that requires knowledge of only the first and second order statistics of the HI, for which we derive closed form expressions.
The proposed estimator reduces to the maximum likelihood estimator (MLE) in the case of ideal hardware.
We also describe FD and HD orthogonal pilot schemes that minimize the mean square error of the MLE in the case of ideal hardware.
We verify the performance of the estimator under varying conditions of transceiver and RIS HI via numerical simulations.
\end{abstract}


\section{Introduction}
Reconfigurable intelligent surface (RIS) technology is an appealing complement to full-duplex (FD) communications because simultaneous transmission occurs naturally through passive reflection without requiring any extra signal processing.
In RIS-assisted communication systems, knowledge of the channel state information (CSI) is fundamental to solving problems on beamforming, self-interference cancellation, and sum rate maximization \cite{Peng2021,Zhao2020,Nguyen2021}.
However, the literature on channel estimation (CE) in RIS-assisted FD systems is limited and only considers special cases of FD transmission \cite{Hu2021,Lin2022,Tekbiyik2021}.
Hu \textit{et al.} considers a multiple-input-multiple-output (MIMO) system and performs CE at the access point (AP) by using two half-duplex (HD) transmission stages: in the first stage the AP transmits and the AP-RIS channel is estimated, while in the second the user equipment (UE) transmits and the UE-RIS channel is estimated \cite{Hu2021}.
A similar approach is adopted in \cite{Lin2022} where only one UE operates in FD while all other UEs and the AP are HD receivers. 
In \cite{Tekbiyik2021}, a graph attention network is used to estimate the cascaded (AP-RIS-UEs) channel  but only considers a single-input-single-output transceiver model.
    
A major design consideration for practical systems is hardware impairments (HI) that appear in cost efficient transceiver implementations and cause a variety of accumulating  errors (e.g., quantization, nonlinear power amplification, I/Q imbalance, and oscillator phase noise) \cite{Papazafeiropoulos2022,Xia2015}.
HI are also present in the RIS, where the controllable reflection coefficient takes on discrete values or is a non-linear function of input voltage which can result in phase errors \cite{Papazafeiropoulos2022,Badiu2020}.
Although the RIS provides increased power gains at the receiver as it has been demonstrated in \cite{Dai2020,Tang2021,Pei2021}, the errors induced by the HI may mitigate these benefits.

It is not surprising that the presence of HI can severely affect the performance of CE algorithms. Indeed, it has been shown to cause error floors in CE accuracy for HD CE methods \cite{Papazafeiropoulos2022}.
In FD systems, the only CE paper to consider HI is Tekbiyik \textit{et al.} \cite{Tekbiyik2021}, however, they only consider receiver HI as random residual loop interference and a deterministic amplitude attenuation imperfection in the RIS coefficient's amplitude (not phase shift offset).

In this letter,
we propose a CE method for RIS MIMO FD systems to jointly estimate all communication channels, i.e., the self-interference, direct, and cascaded channels.
In contrast to \cite{Hu2021,Lin2022,Tekbiyik2021}, our method takes into account the presence of HI on the RIS and on both sides of the communication link. 
Specifically, our main contributions are:
\begin{itemize}
    \item We build upon the works in \cite{Peng2021,Zhao2020,Nguyen2021,Papazafeiropoulos2022,Badiu2020} to derive a comprehensive system model for a narrowband FD RIS MIMO system allowing for simultaneous transmission from the AP and all UEs, while taking into account the HI at the transmitters, receivers, and RIS.
    \item We reformulate our system model to describe the effect of HI solely by additive error terms, and we derive closed form expressions for the first and second order statistics of the HI-induced errors.
    \item We formulate the CE as a linear inverse problem which we solve for two cases: 1) non-ideal hardware with HI assuming HI statistics are known, and 2) ideal hardware.
    \item We propose FD and HD orthogonal pilot schemes based on minimizing the mean square error (MSE) of the least squares (LS) estimator under ideal hardware conditions.
\end{itemize}

\textit{Notation}: Column vectors are denoted as boldface lowercase ($\textbf{a}$), matrices as boldface uppercase ($\textbf{A}$), and scalars as uppercase ($A$) or lowercase ($a$).
Matrix operations on a matrix $\textbf{A}$ are denoted as: conjugate $\textbf{A}^*$, transpose $\textbf{A}^T$, conjugate transpose $\textbf{A}^H$, inverse $\textbf{A}^{-1}$, and trace $\text{Tr}(\textbf{A})$.
A diagonal square matrix with the elements of a vector $\textbf{d}$ on its diagonal is expressed as $\text{diag}(\textbf{d})$. 
The function $\text{vec}(\textbf{A})$ creates a vector by stacking the columns of $\textbf{A}$. 
The identity matrix of dimensions $N \times N$ is $\textbf{I}_N$ and $\textbf{0}_{M \times N}$ is an $M \times N$ matrix of zeros. 
The matrix products are denoted as: Hadamard $\odot$, Kronecker $\otimes$, and Khatri-Rao $\diamond$. 
The floor of $x$ is $\lfloor x \rfloor$, and $\mod$ is the modulo operator.
A circular complex multivariate Gaussian distribution with mean $\boldsymbol{\mu}$ and covariance $\boldsymbol{\Sigma}$ is denoted as $\mathcal{CN}(\boldsymbol{\mu}, \boldsymbol{\Sigma})$.
The von Mises distribution with mean $\mu$ and concentration factor $\kappa$ is denoted as $\mathcal{VM}(\mu, \kappa)$.
Finally, the expectation operator with respect to $\textbf{x}$ is denoted as $\mathbb{E}_\textbf{x}[\cdot]$ or simply $\mathbb{E}[\cdot]$. 

\section{System Model}
\subsection{Ideal System Model}
\begin{figure}[t]
    \centering
    \includegraphics[width=\columnwidth]{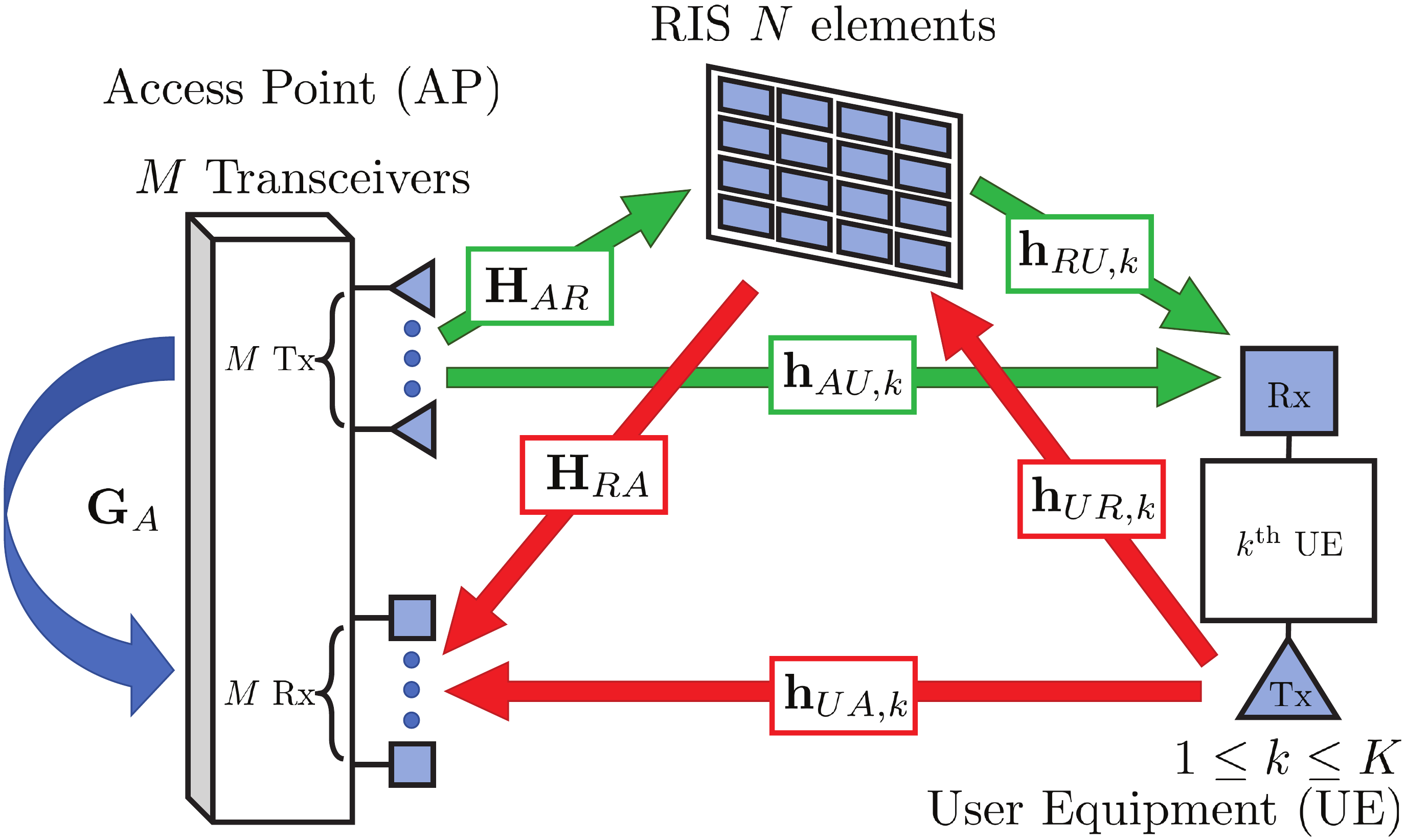}
    \caption{Full duplex MIMO RIS communication model.}
    \label{fig:system-model}
\end{figure}

We consider a narrowband MIMO separate-antenna FD \cite{Sabharwal2014} RIS-assisted communication system comprising an AP with $M$ transmit and $M$ receive antennas, $K$ UEs with one transmit and one receive antenna, and an RIS with $N$ elements as shown in Fig. \ref{fig:system-model}. 
The channels between the (A)ccess Point, (U)ser Equipment, and (R)IS pertinent to this work are defined in Fig.~\ref{fig:system-model}.
For example, $\textbf{H}_{AR}$ is the channel from the AP to the RIS, and $\textbf{G}_A$ is the AP self interference channel. 

Let $\textbf{x}_{A}$ be the transmitted signal from the AP with $\mathbb{E}[\textbf{x}_{A}\textbf{x}_{A}^H] = P_{A} \textbf{I}_M$ where $P_{A}$ is the AP transmit power per antenna. 
Similarly, let $\textbf{x}_{U}$ be the vector containing the transmitted symbols of the $K$ UEs with $\mathbb{E}[\textbf{x}_{U}\textbf{x}_{U}^H] = P_{U} \textbf{I}_K$ where $P_{U}$ is the UEs' transmit power.
The received signal at the AP due to the simultaneous transmission of $\textbf{x}_{A}$ and $\textbf{x}_{U}$ is:
\begin{equation}
\label{eq:idealTx}
    \textbf{y}_{A} = (\textbf{G}_A + \textbf{H}_{RA} \boldsymbol{\Phi} \textbf{H}_{AR})\textbf{x}_{A} + (\textbf{H}_{UA} + \textbf{H}_{RA} \boldsymbol{\Phi} \textbf{H}_{UR}) \textbf{x}_{U} + \textbf{n}_{A}
\end{equation}
where $\textbf{H}_{UA} = [\textbf{h}_{UA,1}, \ldots, \textbf{h}_{UA,K}]$, $\textbf{H}_{UR} =  [\textbf{h}_{UR,1}, \dots, \textbf{h}_{UR,K}]$, and $\textbf{n}_{A} \sim \mathcal{CN}(\textbf{0}, \sigma^2 \textbf{I}_M)$ is the additive white Gaussian noise (AWGN) at the AP.
Finally, $\boldsymbol{\Phi} = \text{diag}(\boldsymbol{\phi}^T) = \text{diag}([e^{j\theta_{1}}, \ldots,  e^{j\theta_{N}}])$  where $\theta_{n} \in [-\pi, \pi)$ is the phase of the $n$th RIS element, $1 \le n \le N$.

The problem we consider in this work is to estimate at the AP\footnote{Estimation at the UEs can be done in a similar manner.} the self-interference channel $\textbf{G}_A$, the direct channel $\textbf{H}_{UA}$, and the cascaded channels from the UEs to the AP and from the AP to the AP through the RIS\footnote{Knowledge of the cascaded channel is sufficient for a variety of processing tasks at the AP including beamforming \cite{Mishra2019}.} in the presence of HI.

\subsection{System Model with Hardware Impairments}
In this section, we build upon \cite{Peng2021,Zhao2020,Nguyen2021,Papazafeiropoulos2022,Badiu2020} to describe the received signal at the AP taking into account HI at the RIS, transmitters, and receivers.

\subsubsection{RIS Hardware Impairments}
With the advancement of lossless metasurfaces \cite{Badloe2017,Epstein2016}, it is possible to design the RIS with maximum signal reflection for the given operating frequency band. 
Meaning HI at the RIS can be assumed to only induce a random offset
$\tilde{\theta}_n \in [-\pi,\pi)$ to the phase of each RIS element, i.e., 
the actual phase of the $n$th element is $\theta_n + \tilde{\theta}_n$ \cite{Papazafeiropoulos2022,Badiu2020}.
All offsets are modelled as i.i.d. von Mises random variables $\tilde{\theta}_n \sim \mathcal{VM}(0, \kappa_{RIS})$ \cite{Papazafeiropoulos2022,Badiu2020}.
The concentration factor $\kappa_{RIS} \ge 0$ represents the severity of the HI, the larger $\kappa_{RIS}$ is, the more concentrated $\tilde{\theta}_n$ is around zero phase offset. 
The case $\kappa_{RIS} = 0$ is the worst case scenario for $\tilde{\theta}_n$, making it a uniform distribution on $[-\pi, \pi)$.

\subsubsection{Transmitter Hardware Impairments}
Transmitter HI are modelled as zero mean additive Gaussian noise \cite{Papazafeiropoulos2022}.
Specifically, the actual transmitted signal by the AP is $\textbf{x}_{A} + \boldsymbol{\delta}_{t_A}$ with $ \boldsymbol{\delta}_{t_A} \sim \mathcal{CN}(\textbf{0}, \boldsymbol{\Sigma}_{t_A}) \in \mathbb{C}^{M \times 1}$ and by the UEs is $\textbf{x}_{U} + \boldsymbol{\delta}_{t_U}$ with $\boldsymbol{\delta}_{t_U} \sim \mathcal{CN}(\textbf{0}, \boldsymbol{\Sigma}_{t_U}) \in \mathbb{C}^{K \times 1}$.
In both cases, the covariance is proportional to the corresponding transmit power, i.e., $\boldsymbol{\Sigma}_{t_A} = \sigma^2_{t_A} P_{A} \textbf{I}_M$ and $\boldsymbol{\Sigma}_{t_U} = \sigma^2_{t_U} P_{U} \textbf{I}_K$ with the constants $\sigma^2_{t_A}$ and $\sigma^2_{t_U}$ measuring the severity of the HI.

\subsubsection{Receiver Hardware Impairments}
The receive signal at the AP with HI is:
\begin{flalign}
    \textbf{y}_{A} & = (\textbf{G}_A + \textbf{H}_{RA} \boldsymbol{\Phi} \tilde{\boldsymbol{\Phi}} \textbf{H}_{AR}) (\textbf{x}_{A} + \boldsymbol{\delta}_{t_A}) \notag \\
    &  \quad + (\textbf{H}_{UA} + \textbf{H}_{RA} \boldsymbol{\Phi} \tilde{\boldsymbol{\Phi}} \textbf{H}_{UR}) (\textbf{x}_{U} + \boldsymbol{\delta}_{t_U}) + 
    \boldsymbol{\delta}_{r_A} + \textbf{n}_A \label{eq:HIyA}
\end{flalign}
where $\tilde{\boldsymbol{\Phi}} = \text{diag}(\tilde{\boldsymbol{\phi}}^T) = \text{diag}([e^{j \tilde{\theta}_1}, \ldots, e^{j \tilde{\theta}_N}])$ contains the HI phase offsets.
In (\ref{eq:HIyA}), $\boldsymbol{\delta}_{r_A}$ denotes the HI at the AP receiver, which is modelled \cite{Xia2015} as zero mean additive Gaussian noise $\boldsymbol{\delta}_{r_A} \sim \mathcal{CN}(\textbf{0}, \boldsymbol{\Sigma}_{r_A}) \in \mathbb{C}^{M \times 1}$.
 Its covariance matrix depends on $\sigma^2_{r_A}$ as $\boldsymbol{\Sigma}_{r_A} = \sigma^2_{r_A} (\textbf{I}_M \odot \boldsymbol{\Gamma})$ where $\boldsymbol{\Gamma} = \mathbb{E}[\textbf{y}_A\textbf{y}_A^H | \boldsymbol{\delta}_{t_A} = \textbf{0}, \boldsymbol{\delta}_{t_U} = \textbf{0}, \boldsymbol{\delta}_{r_A} = \textbf{0}, \textbf{n}_A  = \textbf{0}]$ is the covariance of the received signal with no transmitter or receiver HI:
\begin{align}
    &\boldsymbol{\Gamma} = P_{A} ( \textbf{G}_A\textbf{G}_A^H + \varphi \textbf{G}_A\textbf{H}_{AR}^H \boldsymbol{\Phi}^H \textbf{H}_{RA}^H + \varphi \textbf{H}_{RA} \boldsymbol{\Phi} \textbf{H}_{AR} \textbf{G}_A^H \notag \\
    &\quad \quad + \textbf{H}_{RA} \boldsymbol{\Phi} (\varphi^2 \textbf{H}_{AR} \textbf{H}_{AR}^H + (1-\varphi^2)\textbf{I}_N) \boldsymbol{\Phi}^H \textbf{H}_{RA}^H ) \notag \\
    &+ P_{U} ( \textbf{H}_{UA}\textbf{G}_A^H + \varphi \textbf{H}_{UA}\textbf{H}_{UR}^H \boldsymbol{\Phi}^H \textbf{H}_{RA}^H + \varphi \textbf{H}_{RA} \boldsymbol{\Phi} \textbf{H}_{UR} \textbf{H}_{UA}^H \notag \\
    &\quad \quad + \textbf{H}_{RA} \boldsymbol{\Phi} (\varphi^2 \textbf{H}_{UR} \textbf{H}_{UR}^H + (1-\varphi^2)\textbf{I}_N) \boldsymbol{\Phi}^H \textbf{H}_{RA}^H) \label{eq:hi-EyAyAH}
\end{align}
In (\ref{eq:hi-EyAyAH}),  $\varphi = \mathbb{E}[e^{j\tilde{\theta}_n}]  = \frac{I_1(\kappa_{RIS})}{I_0(\kappa_{RIS})} \in [0,1]$ with $I_p(\cdot)$ being the modified Bessel function of the first kind of order $p$.
This is an extension from HD to FD based on \cite{Papazafeiropoulos2022} assuming the AP and UE transmit signals are independent $\mathbb{E}[\textbf{x}_A\textbf{x}_U^H] = \textbf{0}$.

As a final note, all HI random variables $\boldsymbol{\delta}_{t_A}$, $\boldsymbol{\delta}_{t_U}$, $\boldsymbol{\delta}_{r_A}$, and $\tilde{\theta}_n$ are independent to each other.

\section{Channel Estimation}
To estimate the CSI at the AP, the AP and UEs transmit pilot symbols over the training period of length $T$.
The $t$th receive vector at the AP due to the $t$th pilot symbol transmitted from the AP, $\textbf{x}_{A,t}$, and UEs, $\textbf{x}_{U,t}$, is:
\begin{flalign}
    \textbf{y}_{A,t} & = (\textbf{G}_A + \textbf{H}_{RA} \boldsymbol{\Phi}_t \tilde{\boldsymbol{\Phi}}_t \textbf{H}_{AR}) (\textbf{x}_{A,t} + \boldsymbol{\delta}_{t_A,t}) \notag \\
    & \quad + (\textbf{H}_{UA} + \textbf{H}_{RA} \boldsymbol{\Phi}_t \tilde{\boldsymbol{\Phi}}_t \textbf{H}_{UR}) (\textbf{x}_{U,t} + \boldsymbol{\delta}_{t_U,t}) \notag \\
    & \quad + \boldsymbol{\delta}_{r_A,t} + \textbf{n}_{A,t}, \,\,\,\, t\in\{1, \ldots, T\} \label{eq:HIyAt}
\end{flalign}
where $\boldsymbol{\Phi}_t = \text{diag}(\boldsymbol{\phi}_t)$ are the RIS phase coefficients during the $t$th pilot transmission.
 
Using vectorization, Kronecker, and Khatri-Rao products similar to HD systems in \cite{Swindlehurst2022}, we can rewrite (\ref{eq:HIyAt}) as:
\begin{equation}
\label{eq:HIyAtZth}
    \textbf{y}_{A,t} =
        \textbf{Z}_t
        \textbf{h}
        + \boldsymbol{\delta}_{r_A,t} + \textbf{n}_{A,t}
\end{equation}
where $\textbf{h} \in \mathbb{C}^{M(M+K)(N+1) \times 1}$ gathers the channels as follows:
\begin{equation}
\label{eq:h}
    \textbf{h} = \begin{bmatrix} 
        \text{vec}(\textbf{G}_A) \\ 
        \text{vec}(\textbf{H}_{AR}^T \diamond \textbf{H}_{RA}) \\ 
        \text{vec}(\textbf{H}_{UA}) \\ 
        \text{vec}(\textbf{H}_{UR}^T \diamond \textbf{H}_{RA})
        \end{bmatrix}
\end{equation}
and $\textbf{Z}_t \in \mathbb{C}^{M \times M(M+K)(N+1)}$ is given by
\begin{equation}
\label{eq:Zt}
    \textbf{Z}_t = \begin{bmatrix}
        \textbf{x}_{A,t} + \boldsymbol{\delta}_{t_A,t} \\
        (\boldsymbol{\phi}_t \odot \tilde{\boldsymbol{\phi}}_t) \otimes (\textbf{x}_{A,t} + \boldsymbol{\delta}_{t_A,t}) \\
        \textbf{x}_{U,t} + \boldsymbol{\delta}_{t_U,t} \\
        (\boldsymbol{\phi}_t \odot \tilde{\boldsymbol{\phi}}_t) \otimes (\textbf{x}_{U,t} + \boldsymbol{\delta}_{t_U,t}) \\
        \end{bmatrix}^T \otimes \textbf{I}_M
\end{equation}

Since $\textbf{Z}_t$ contains multiplicative and additive random errors, we reformulate (\ref{eq:Zt}) to contain only additive errors through a similar procedure as in \cite{Nicholson2020}. Specifically, we can write $\textbf{Z}_t = (\textbf{x}_t + \textbf{e}_t)^T \otimes \textbf{I}_M$, where $\textbf{x}_t, \textbf{e}_t \in \mathbb{C}^{(M+K)(N+1) \times 1}$ are:
\begin{align}
    \textbf{x}_t &= \label{eq:x_t}
        \begin{bmatrix}
        \textbf{x}_{A,t}^T &
        \boldsymbol{\phi}_t^T \otimes \textbf{x}_{A,t}^T &
        \textbf{x}_{U,t}^T &
        \boldsymbol{\phi}_t^T \otimes \textbf{x}_{U,t}^T
        \end{bmatrix}^T \\
    \textbf{e}_t &= \label{eq:e_t}
        \begin{bmatrix}
        \boldsymbol{\delta}_{t_A,t} \\
        (\boldsymbol{\phi}_t \odot \tilde{\boldsymbol{\phi}}_t - \boldsymbol{\phi}_t) \otimes \textbf{x}_{A,t} + (\boldsymbol{\phi}_t \odot \tilde{\boldsymbol{\phi}}_t) \otimes \boldsymbol{\delta}_{t_A,t} \\
        \boldsymbol{\delta}_{t_U,t} \\
        (\boldsymbol{\phi}_t \odot \tilde{\boldsymbol{\phi}}_t - \boldsymbol{\phi}_t) \otimes \textbf{x}_{U,t} + (\boldsymbol{\phi}_t \odot \tilde{\boldsymbol{\phi}}_t) \otimes \boldsymbol{\delta}_{t_U,t} \\
        \end{bmatrix}
\end{align}

We can obtain the statistics of this additive error $\textbf{e}_t$ in terms of known values. The mean is given by
\begin{equation}
\label{eq:E_et}
    \mathbb{E}[\textbf{e}_t] =
    \begin{bmatrix}
    \textbf{0}_{M \times 1} \\
    ((\varphi - 1) \boldsymbol{\phi}_t ) \otimes \textbf{x}_{A,t} \\
    \textbf{0}_{K \times 1} \\
    ((\varphi - 1) \boldsymbol{\phi}_t ) \otimes \textbf{x}_{U,t} \\
    \end{bmatrix}
\end{equation}
and the correlation matrix is given by
\begin{flalign}
    & \mathbb{E}[\textbf{e}_t\textbf{e}_t^H] = \label{eq:E_etetH} \\
    & \begin{bmatrix}
        \boldsymbol{\Sigma}_{t_A} & \varphi \boldsymbol{\phi}_t^H \otimes \boldsymbol{\Sigma}_{t_A} & \textbf{0}_{M \times K} & \textbf{0}_{M \times KN} \\
        \varphi \boldsymbol{\phi}_t \otimes \boldsymbol{\Sigma}_{t_A} & \textbf{A}_t & \textbf{0}_{MN \times K} & \textbf{B}_t \\
        \textbf{0}_{K \times M} & \textbf{0}_{K \times MN} & \boldsymbol{\Sigma}_{t_U} & \varphi \boldsymbol{\phi}_t^H \otimes \boldsymbol{\Sigma}_{t_U} \\
        \textbf{0}_{KN \times M} & \textbf{C}_t & \varphi \boldsymbol{\phi}_t \otimes \boldsymbol{\Sigma}_{t_U} & \textbf{D}_t
    \end{bmatrix} \notag
\end{flalign}
\begin{align}
    \textbf{A}_t &= ((\varphi^2 - 2\varphi + 1) \boldsymbol{\phi}_t\boldsymbol{\phi}_t^H + (1 - \varphi^2) \textbf{I}_N) \otimes \textbf{x}_{A,t} \textbf{x}_{A,t}^H \notag \\
    & \quad + (\varphi^2 \boldsymbol{\phi}_t\boldsymbol{\phi}_t^H + (1-\varphi^2)\textbf{I}_N) \otimes \boldsymbol{\Sigma}_{t_A} \label{eq:At} \\
    \textbf{B}_t & = ((\varphi^2 - 2\varphi + 1) \boldsymbol{\phi}_t\boldsymbol{\phi}_t^H + (1 - \varphi^2) \textbf{I}_N) \otimes \textbf{x}_{A,t} \textbf{x}_{U,t}^H \label{eq:Bt} \\
    \textbf{C}_t & = ((\varphi^2 - 2\varphi + 1) \boldsymbol{\phi}_t\boldsymbol{\phi}_t^H + (1 - \varphi^2) \textbf{I}_N) \otimes \textbf{x}_{U,t} \textbf{x}_{A,t}^H \label{eq:Ct} \\
    \textbf{D}_t & = ((\varphi^2 - 2\varphi + 1) \boldsymbol{\phi}_t\boldsymbol{\phi}_t^H + (1 - \varphi^2) \textbf{I}_N) \otimes \textbf{x}_{U,t} \textbf{x}_{U,t}^H \notag \\
    & \quad + (\varphi^2 \boldsymbol{\phi}_t\boldsymbol{\phi}_t^H + (1-\varphi^2)\textbf{I}_N) \otimes \boldsymbol{\Sigma}_{t_U} \label{eq:Dt} 
\end{align}
The complete derivation of (\ref{eq:E_etetH}) is omitted due to lack of space but can be proved using (\ref{eq:e_t}) and the HI statistics.

We note that being able to rewrite $\textbf{Z}_t$ in terms of only additive errors gives insight into how the actual values deviate away from the desired values.
We see from (\ref{eq:E_et}) that $\textbf{e}_t$ would only have zero mean if there were no HI in the RIS.
While (\ref{eq:E_etetH}) shows that all HI of transmitters, receivers, and RIS contribute to the correlation of error. 

Based on the first and second order statistics, the additive HI error depends on pilot symbols and the RIS phase shifts. 

\subsection{Linear Inverse Problem}
We stack the $T$ received vectors at the AP into a column vector $\textbf{\=y}_A =
[\textbf{y}_{A,1}^T, \ldots, \textbf{y}_{A,T}^T]^T \in \mathbb{C}^{TM \times 1}$ as:
\begin{equation}
\label{eq:gammaA}
    \textbf{\=y}_A
    =
    (\textbf{X} + \textbf{E})
    \textbf{h}  +
    \boldsymbol{\Delta}_{r_A} +
    \textbf{\=n}_{A}
\end{equation}
where $\textbf{X} = [\textbf{x}_1, \ldots, \textbf{x}_T]^T \otimes \textbf{I}_M$,  $\textbf{E} = [\textbf{e}_1, \ldots, \textbf{e}_T]^T \otimes \textbf{I}_M$, $\boldsymbol{\Delta}_{r_A} =
[\boldsymbol{\delta}_{r_A,1}^T, \ldots, \boldsymbol{\delta}_{r_A,T}^T
]^T$, and $\textbf{\=n}_{A} =
[\textbf{n}_{A,1}^T, \ldots, \textbf{n}_{A,T}^T
]^T$.
The matrix $\textbf{X} \in \mathbb{C}^{TM \times M(M+K)(N+1)}$ contains the pilots and RIS phase shifts over the training period and $\textbf{E}$ contains the corresponding additive HI errors.

We propose to estimate ${\textbf{h}}$ by minimizing the expected value with respect to $\textbf{E}$ of the squared error:
\begin{equation}
\label{eq:min_with_E}
\min_{\textbf{h}} \mathbb{E}_\textbf{E}[||\textbf{\=y}_A - (\textbf{X} + \textbf{E}) \textbf{h}||_2^2]
\end{equation}
which yields the following estimate 
\begin{flalign}
& \hat{\textbf{h}}_{HI} = \label{eq:h_HI} \\
& (\textbf{X}^H\textbf{X} +
\textbf{X}^H \mathbb{E}[\textbf{E}] +
\mathbb{E}[\textbf{E}^H] \textbf{X}  + 
\mathbb{E}[\textbf{E}^H\textbf{E}])^{-1}
(\textbf{X} + \mathbb{E}[\textbf{E}])^H
\textbf{\=y}_A \notag
\end{flalign}
where $\mathbb{E}[\textbf{E}] = [\mathbb{E}[\textbf{e}_1], \ldots, \mathbb{E}[\textbf{e}_T]]^T \otimes \textbf{I}_M$ and $\mathbb{E}[\textbf{E}^H\textbf{E}] = \Sigma_{t=1}^T \mathbb{E}[\textbf{e}_t\textbf{e}_t^H]^*\otimes \textbf{I}_M$.
It can be shown that $\hat{\textbf{h}}_{HI}$ is an unbiased estimator with MSE:
\begin{flalign}
\text{MSE}_{HI} =
& \text{Tr} ( (\textbf{X}^H\textbf{X} +
\textbf{X}^H \mathbb{E}[\textbf{E}] +
\mathbb{E}[\textbf{E}^H] \textbf{X}  + 
\mathbb{E}[\textbf{E}^H\textbf{E}])^{-2} \notag \\
& \quad \ (\textbf{X} + \mathbb{E}[\textbf{E}])^H
\mathbb{E}[\textbf{\=y}_A \textbf{\=y}_A^H] (\textbf{X} + \mathbb{E}[\textbf{E}]) ) \label{eq:MSE_HI}
\end{flalign}
where $\mathbb{E}[\textbf{\=y}_A \textbf{\=y}_A^H] = \textbf{X}\textbf{h}\textbf{h}^H\textbf{X}^H + \textbf{X}\textbf{h}\textbf{h}^H\mathbb{E}[\textbf{E}^H] + \mathbb{E}[\textbf{E}]\textbf{h}\textbf{h}^H\textbf{X}^H + \mathbb{E}[\textbf{E}\textbf{h}\textbf{h}^H\textbf{E}^H] + \textbf{I}_T \otimes \boldsymbol{\Sigma}_{r_A} + \sigma^2 \textbf{I}_{TM}$.

When there are no HI, (\ref{eq:min_with_E}) reduces to:
\begin{equation}
\label{eq:min_without_E}
\min_{\textbf{h}} ||\textbf{\=y}_A - \textbf{X} \textbf{h}||_2^2
\end{equation}
which leads to the LS solution which is also the maximum likelihood estimate (MLE) of $\textbf{h}$:
\begin{equation}
\label{eq:LS}
\hat{\textbf{h}}_{LS} = 
    (\textbf{X}^H\textbf{X})^{-1}\textbf{X}^H
    \textbf{\=y}_A
\end{equation}
The MSE of $\hat{\textbf{h}}_{LS}$ when there are no HI is:
\begin{equation}
    \label{eq:MSE_LS}
    \text{MSE}_{LS} = \sigma^2 \text{Tr} ( (\textbf{X}^H\textbf{X})^{-1} )
\end{equation}
where this estimator is efficient and attains the Cram\'er Rao lower bound (CRLB) under ideal hardware conditions \cite[p. 529-530]{Kay1993}.
To the best of our knowledge, (\ref{eq:h_HI}) and (\ref{eq:LS}) are the first channel estimates reported for RIS-assisted FD MIMO with ideal and non-ideal hardware, respectively, that allow for simultaneous transmission from both the AP and all UEs.

\subsection{Training scheme: Choice of pilots and RIS phase values}
\label{sec:Pilots}
Ideally the pilots and RIS phase values should be chosen to  minimize the MSE of the proposed estimator $\hat{\textbf{h}}_{HI}$. However, this is difficult to do considering (\ref{eq:MSE_HI}) depends on the channels being estimated. 
Instead, we propose to choose the pilots and RIS phase values to minimize the MSE of the LS estimator $\hat{\textbf{h}}_{LS}$, which corresponds to minimizing the CRLB under ideal hardware conditions.
To minimize (\ref{eq:MSE_LS}), we should choose $\textbf{X}$ such that the positive semi-definite term $\textbf{X}^H\textbf{X} = \sum_{t=1}^T (\textbf{x}_t \textbf{x}_t^H)^* \otimes \textbf{I}_M$ is diagonal and has maximum trace (see \cite[Ex 3.12]{Jensen2020,Kay1993}). This has the added benefit of reducing the complexity of the matrix inversion in (\ref{eq:h_HI}) and (\ref{eq:LS}) from $O(n^3)$ to $O(n)$ where $n$ is the dimension of  $\textbf{X}^H\textbf{X}$.

To that end, we see from (\ref{eq:x_t}) that $\boldsymbol{\phi}_t$, $\textbf{x}_{A,t}$, $\textbf{x}_{U,t}$, $\boldsymbol{\phi}_t$, $\textbf{x}_{A,t}$, and $\textbf{x}_{U,t}$
must be chosen such that: (i) $\Sigma_{t=1}^T \boldsymbol{\phi}_t \boldsymbol{\phi}_t^H = T\textbf{I}_N$, (ii) $\Sigma_{t=1}^T \boldsymbol{\phi}_t = \textbf{0}_{N \times 1}$, (iii) $\Sigma_{t=1}^T \textbf{x}_{A,t}\textbf{x}_{A,t}^H = \text{diag}([\mathcal{E}_{A,1}, \ldots, \mathcal{E}_{A,M}])$, (iv) $\Sigma_{t=1}^T \textbf{x}_{U,t} \textbf{x}_{U,t}^H = \text{diag}([\mathcal{E}_{U,1}, \ldots, \mathcal{E}_{U,K}])$, and (v) $\Sigma_{t=1}^T \textbf{x}_{A,t}\textbf{x}_{U,t}^H = \textbf{0}_{M \times K}$.
The last condition implies that $\textbf{x}_{A,t}$ and $\textbf{x}_{U,t}$ are orthogonal.

To meet conditions (i) and (ii), we adopt the popular RIS DFT method \cite{Jensen2020} where the training period is partitioned into $N+1$ blocks composed of $L$ pilots each, i.e., $L = T/(N+1)$, where $L$ is assumed to be an integer.
During each block $b\in\{1, \ldots, N+1\}$, the RIS phase values are set to $\boldsymbol{\phi}_t = \boldsymbol{\phi}_b$, $t \in \{(b-1)L+1, \ldots, bL\}$. 
From one block to the next, the RIS phase values are updated such that $\Bar{\boldsymbol{\phi}}_b = [1,\boldsymbol{\phi}_b^T]^T$ cycles through all column vectors of an ($N+1$)-sized DFT matrix.
Within each block, the AP and UEs transmit $L$ pilots cycling over all column vectors of $\textbf{S}_A \in \mathbb{C}^{M \times L}$ and $\textbf{S}_U \in \mathbb{C}^{K \times L}$, respectively.
All pilot transmissions from the AP and the UEs are then given by $[\textbf{x}_{A,1}, \ldots, \textbf{x}_{A,T}] = \sqrt{P_A}[\textbf{S}_{A}, \ldots, \textbf{S}_{A}] \in \mathbb{C}^{M \times T}$ and $[\textbf{x}_{U,1}, \ldots, \textbf{x}_{U,T}] = \sqrt{P_U}[\textbf{S}_{U}, \ldots, \textbf{S}_{U}] \in \mathbb{C}^{K \times T}$.
To obtain a unique estimate of $\textbf{h}$, $\textbf{X}$ must be full column rank and we need a pilot training overhead of $T \ge (M+K)(N+1)$.

There are several ways to construct $\textbf{S}_A$ and $\textbf{S}_U$ to meet conditions (iii)-(v). We present three such constructions in Table \ref{tab:pilotschemes} when $M \ge K$. In this table,  $\textbf{P}$ is given by
\begin{align}
\textbf{P} &= 
\begin{bmatrix}\smash[b]{\underbrace{\begin{matrix}\textbf{Q}_K, & 
\ldots, & 
\textbf{Q}_K\end{matrix}}_{q\text{ times}}}, & 
\begin{bmatrix}
    \textbf{Q}_r \\
    \textbf{0}_{(K-r) \times r}
\end{bmatrix}
\end{bmatrix} \in \mathbb{C}^{K \times M}
\end{align}
with $q = \lfloor \frac{M}{K} \rfloor$, $r = M \mod K$, and $\textbf{Q}_M$, $\textbf{Q}_K$ and $\textbf{Q}_r$ are orthonormal matrices of size $M$, $K$, and $r$, respectively, e.g., normalized DFT matrices.
It is easy to check that $\textbf{S}_{A}\textbf{S}_{A}^H$ and $\textbf{S}_{U}\textbf{S}_{U}^H$ are diagonal matrices, and $\textbf{S}_{A}\textbf{S}_{U}^H = \textbf{0}_{M \times K}$, satisfying orthogonality conditions (iii)-(v).
In Table \ref{tab:pilotschemes}, Scheme 1 operates in FD while Schemes 2 and 3 operate in HD, meaning only scheme 1 allows for the simultaneous transmission of AP and UE pilots.
Scheme 3 has the minimum value of $T$ required to ensure $\textbf{X}$ is full rank.
To satisfy conditions (iii)-(v) with FD transmission, Scheme 1 has a slightly higher pilot overhead.
Scheme 2 has a similar pilot structure to Scheme 1 with the same pilot overhead, except it operates in HD.

\begin{table}[htbp]
\caption{Proposed orthogonal pilot transmission schemes}
\begin{center}
\begin{tabular}{@{}llll@{}}\toprule
Scheme & 1 & 2 & 3 \\ \midrule
Duplex & FD & HD & HD \\ 
$\textbf{S}_A$ & $[\textbf{Q}_M, \textbf{Q}_M]$ & $[\textbf{Q}_M, \textbf{0}_{M \times M}]$ & $[\textbf{Q}_M, \textbf{0}_{M \times K}]$ \\ 
$\textbf{S}_U$ & $[\textbf{P}, -\textbf{P}]$ & $[\textbf{0}_{K \times M}, \textbf{P}]$ & $[\textbf{0}_{K \times M}, \textbf{Q}_K]$ \\ 
$T$ & $2M(N+1)$ & $2M(N+1)$ & $(M+K)(N+1)$ \\ \bottomrule
\end{tabular}
\label{tab:pilotschemes}
\end{center}
\end{table}

\section{Numerical Results}

\begin{figure*}[t]
\centering
\begin{subfigure}[b]{0.325\textwidth}
    \centering
    \includegraphics[width=\textwidth]{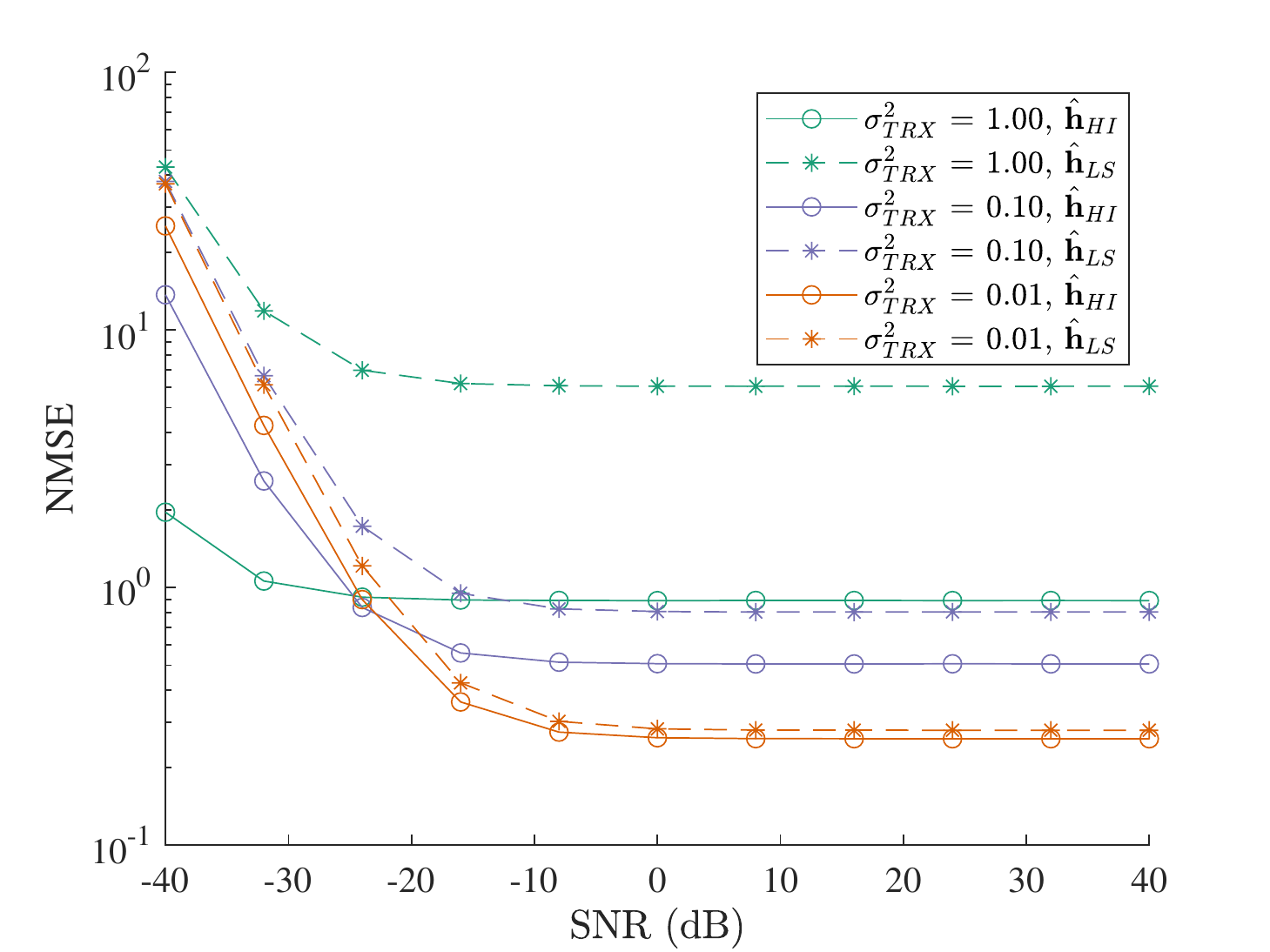}
    \caption{}
    \label{fig:NMSE_vs_SNR}
\end{subfigure} \hfill
\begin{subfigure}[b]{0.325\textwidth}
    \centering
    \includegraphics[width=\textwidth]{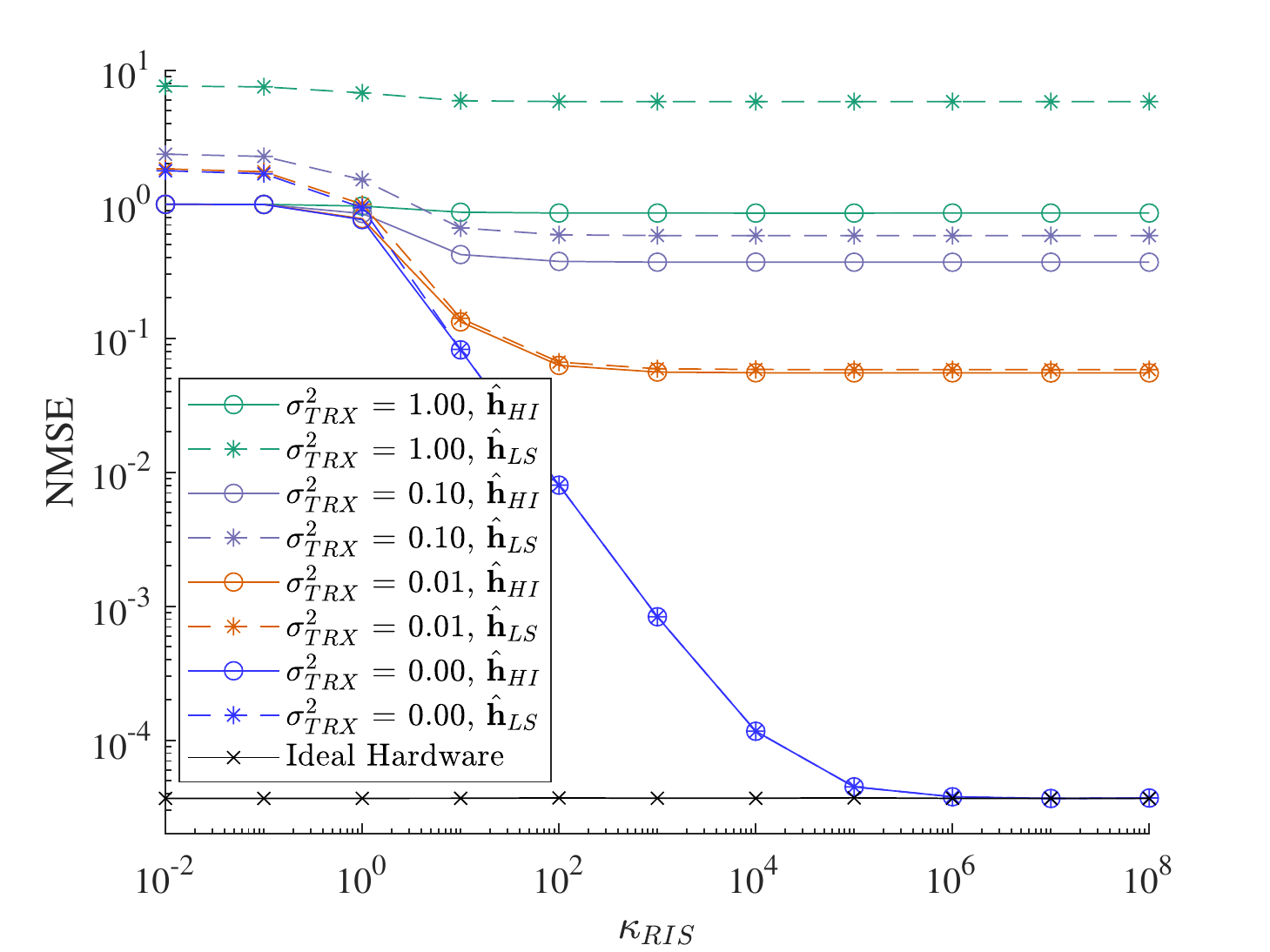}
    \caption{}
    \label{fig:NMSE_vs_kris}
\end{subfigure} \hfill
\begin{subfigure}[b]{0.325\textwidth}
    \centering
    \includegraphics[width=\textwidth]{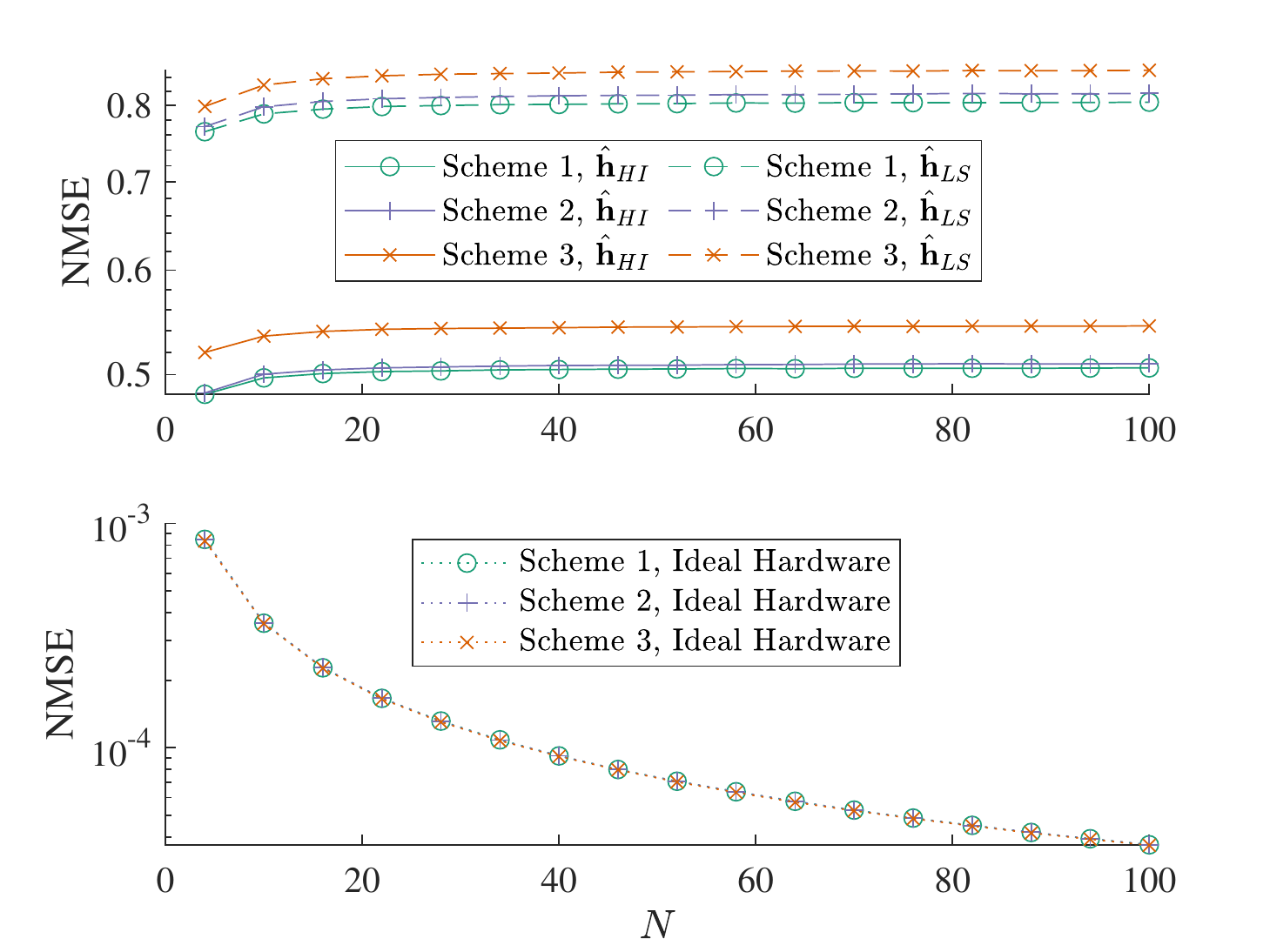}
    \caption{}
    \label{fig:NMSE_vs_N}
\end{subfigure}
\caption{NMSEs of the least squares ($\hat{\textbf{h}}_{LS}$) and the proposed ($\hat{\textbf{h}}_{HI}$) estimators vs: (a) SNR, (b) $\kappa_{RIS}$, and (c) $N$.}
\label{fig:NMSE}
\end{figure*}

Unless stated otherwise, we simulate a system with $M=5$ transmit and receive antennas at the AP, $K=2$ single-antenna UEs, and $N=100$ elements at the RIS. 
All channels are modeled as uncorrelated Rayleigh fading with large-scale fading parameters and distances adopted from \cite{Hu2021}. 
Specifically, the AP-RIS distance is 20m with path loss exponent (PLE) of 2.1, UE-RIS distances are 20m with PLE of 4.2, and AP-UE distance is 30m with PLE of 2.2. 
The self-interference channel is modeled as Rayleigh fading having normalized channel power \cite{Lin2022}.
The signal-to-noise ratio (SNR) is defined as $P/\sigma^2$ with $P_A = P_U = P$.
CE performance is measured by the normalized MSE (NMSE) defined as
$\text{NMSE} = ||\textbf{h} - \hat{\textbf{h}}||_2^2 / ||\textbf{h}||_2^2$.
All results shown are obtained by Monte-Carlo simulations over 10000 independent channel realizations. 

We use a non-ideal hardware communication system to compare the performance of the proposed estimator in (\ref{eq:h_HI}) that uses knowledge of HI statistics to a naive LS estimator based on (\ref{eq:LS}) that does not take into account any HI.
Fig. \ref{fig:NMSE_vs_SNR} shows the NMSE versus SNR using pilot Scheme 1 for different severity levels of transceiver HI.
The RIS HI concentration factor is set to $\kappa_{RIS}=4$ to describe a faulty RIS system \cite{Papazafeiropoulos2022,Badiu2020}. 
Results for different values (1, 0.1, and 0.01) of transceiver HI severity $\sigma^2_{t_A}=\sigma^2_{t_U}=\sigma^2_{r_A}=\sigma^2_{TRX}$, corresponding to 100\%, 10\% and 1\% of the transmit/receive signal power, are included. 
For all considered levels of HI, the NMSE reaches an error floor for SNR values larger than 0dB. 
This indicates that the main source of CE error is due to HI.
The proposed estimator always outperforms the naive LS method throughout the considered SNR range, reaching a significantly lower NMSE error floor.

Fig. \ref{fig:NMSE_vs_kris} shows the NMSE versus RIS HI severity $\kappa_{RIS}$ using pilot Scheme 1 at a SNR of 20 dB which implies that HI is the main source of error as discussed earlier.  
The range of RIS HI $\kappa_{RIS}$ represents phase shift errors going from being nearly uniformly distributed (low values of $\kappa_{RIS}$) to highly concentrated around zero (high values of $\kappa_{RIS}$). 
We include results for different values (1, 0.1, and 0.01) of transceiver HI severity, the case of no transceiver HI, and ideal hardware for reference.
We see that the NMSE is significantly higher than in the case of ideal hardware even when the transceiver HI is 1\% of the transmit power.
Similarly, for zero transceiver HI, the NMSE is significantly high for moderate values of $\kappa_{RIS}$.

Fig. \ref{fig:NMSE_vs_N} shows the NMSE versus number of RIS elements $N$ using all three pilot schemes in Table \ref{tab:pilotschemes}, at a SNR of 20 dB for non-ideal hardware with  $\kappa_{RIS} = 4$ and $\sigma^2_{TRX}=0.1$ (top figure) and ideal hardware (bottom figure).
For a fair comparison, the transmission power for each pilot scheme is adjusted such that the total energy cost at the AP $\mathcal{E}_A = \sum_{m=1}^{M} \mathcal{E}_{A,m} = P_A(N+1)\text{Tr}(\textbf{S}_A\textbf{S}_A^H)$ and at the UEs $\mathcal{E}_U = \sum_{k=1}^{K} \mathcal{E}_{U,k} = P_U(N+1)\text{Tr}(\textbf{S}_U\textbf{S}_U^H)$ are the same for all schemes.
For our setup, if $P^{(1)}_A=P^{(1)}_U=1$ are the powers for Scheme 1, then Scheme 2 transmits at increased powers $P^{(2)}_A=P^{(2)}_U=2$, and Scheme 3 at $P^{(3)}_A=2$ and $P^{(3)}_U=2\frac{M}{K}=5$.
The SNR is now defined as $P^{(1)}/\sigma^2$ with $P^{(1)}_A = P^{(1)}_U = P^{(1)}$.
Pilot schemes 1 and 2 have a pilot training overhead of $T = 10(N+1)$ while scheme 3 has $T = 7(N+1)$.
As the number of RIS elements $N$ increases, the NMSE for the scenario of ideal hardware decreases but increases for non-ideal hardware indicating more faulty RIS elements will not improve the NMSE.
The NMSE results are the same for each scheme under ideal hardware conditions, while under non-ideal hardware, Scheme 1 always has the lowest NMSE, indicating the FD scheme is more robust to HI than the HD schemes.

\section{Conclusion}
We proposed a CE method for narrowband RIS-assisted FD MIMO communication systems to estimate all CSI including the self-interference channel while in the presence of HI at the transmitters, receivers, and RIS elements.
We studied the impact that HI have on the system by formulating them in terms of additive errors, and derived closed-form expressions of the first and second order statistics of the HI.
Assuming knowledge of the HI statistics, our proposed channel estimator is unbiased, and reduces to the MLE in ideal hardware conditions.
We further proposed FD and HD pilot schemes that minimize the MSE of the MLE.
Through simulations, we showed that the proposed estimator has significantly better accuracy than the naive LS estimator, but HI will result in error floors regardless of antenna power gains achieved at the receiver.
Results showed that a FD pilot scheme was more robust to the impact of HI on CE accuracy than HD pilot schemes, but HD may still be of interest for minimum pilot training overhead.

\newpage
\onecolumn
{\appendices
\section{Derivation of the correlation matrix}
The complete expansion of (\ref{eq:E_etetH}) is as follows:

$$
\mathbb{E}[\textbf{e}_t\textbf{e}_t^H] = \mathbb{E}\left[
\begin{bmatrix}
    \boldsymbol{\delta}_{t_A,t} \\
    (\boldsymbol{\phi}_t \odot \tilde{\boldsymbol{\phi}}_t - \boldsymbol{\phi}_t) \otimes \textbf{x}_{A,t} + (\boldsymbol{\phi}_t \odot \tilde{\boldsymbol{\phi}}_t) \otimes \boldsymbol{\delta}_{t_A,t} \\
    \boldsymbol{\delta}_{t_U,t} \\
    (\boldsymbol{\phi}_t \odot \tilde{\boldsymbol{\phi}}_t - \boldsymbol{\phi}_t) \otimes \textbf{x}_{U,t} + (\boldsymbol{\phi}_t \odot \tilde{\boldsymbol{\phi}}_t) \otimes \boldsymbol{\delta}_{t_U,t} \\
\end{bmatrix}
\begin{bmatrix}
    \boldsymbol{\delta}_{t_A,t} \\
    (\boldsymbol{\phi}_t \odot \tilde{\boldsymbol{\phi}}_t - \boldsymbol{\phi}_t) \otimes \textbf{x}_{A,t} + (\boldsymbol{\phi}_t \odot \tilde{\boldsymbol{\phi}}_t) \otimes \boldsymbol{\delta}_{t_A,t} \\
    \boldsymbol{\delta}_{t_U,t} \\
    (\boldsymbol{\phi}_t \odot \tilde{\boldsymbol{\phi}}_t - \boldsymbol{\phi}_t) \otimes \textbf{x}_{U,t} + (\boldsymbol{\phi}_t \odot \tilde{\boldsymbol{\phi}}_t) \otimes \boldsymbol{\delta}_{t_U,t} \\
\end{bmatrix}^H
\right] \\
$$

{\tabulinesep=1.2mm
\begin{table}[ht]
\centering
$$= \mathbb{E}\left[
\begin{tabu}{cccc}
\boldsymbol{\delta}_{t_A,t} \boldsymbol{\delta}_{t_A,t}^H & 
\begin{smallmatrix} \boldsymbol{\delta}_{t_A,t} ((\boldsymbol{\phi}_t^H \odot \tilde{\boldsymbol{\phi}}_t^H - \boldsymbol{\phi}_t^H) \otimes \textbf{x}_{A,t}^H) \\ + \boldsymbol{\delta}_{t_A,t} ((\boldsymbol{\phi}_t^H \odot \tilde{\boldsymbol{\phi}}_t^H) \otimes \boldsymbol{\delta}_{t_A,t}^H) \end{smallmatrix} & 
\boldsymbol{\delta}_{t_A,t} \boldsymbol{\delta}_{t_U,t}^H & 
\begin{smallmatrix} \boldsymbol{\delta}_{t_A,t} ((\boldsymbol{\phi}_t^H \odot \tilde{\boldsymbol{\phi}}_t^H - \boldsymbol{\phi}_t^H) \otimes \textbf{x}_{U,t}^H) \\ + \boldsymbol{\delta}_{t_A,t} ((\boldsymbol{\phi}_t^H \odot \tilde{\boldsymbol{\phi}}_t^H) \otimes \boldsymbol{\delta}_{t_U,t}^H) \end{smallmatrix} \\ 
\begin{smallmatrix} ((\boldsymbol{\phi}_t \odot \tilde{\boldsymbol{\phi}}_t - \boldsymbol{\phi}_t) \otimes \textbf{x}_{A,t})\boldsymbol{\delta}_{t_A,t}^H \\ + ((\boldsymbol{\phi}_t \odot \tilde{\boldsymbol{\phi}}_t) \otimes \boldsymbol{\delta}_{t_A,t}) \boldsymbol{\delta}_{t_A,t}^H \end{smallmatrix} &
\begin{smallmatrix} (\boldsymbol{\phi}_t \odot \tilde{\boldsymbol{\phi}}_t - \boldsymbol{\phi}_t) (\boldsymbol{\phi}_t^H \odot \tilde{\boldsymbol{\phi}}_t^H - \boldsymbol{\phi}_t^H) \otimes \textbf{x}_{A,t} \textbf{x}_{A,t}^H \\ + (\boldsymbol{\phi}_t \odot \tilde{\boldsymbol{\phi}}_t - \boldsymbol{\phi}_t) (\boldsymbol{\phi}_t^H \odot \tilde{\boldsymbol{\phi}}_t^H) \otimes \textbf{x}_{A,t} \boldsymbol{\delta}_{t_A,t}^H \\ + (\boldsymbol{\phi}_t \odot \tilde{\boldsymbol{\phi}}_t) (\boldsymbol{\phi}_t^H \odot \tilde{\boldsymbol{\phi}}_t^H - \boldsymbol{\phi}_t^H) \otimes \boldsymbol{\delta}_{t_A,t} \textbf{x}_{A,t}^H \\ + (\boldsymbol{\phi}_t \odot \tilde{\boldsymbol{\phi}}_t) (\boldsymbol{\phi}_t^H \odot \tilde{\boldsymbol{\phi}}_t^H) \otimes \boldsymbol{\delta}_{t_A,t} \boldsymbol{\delta}_{t_A,t}^H \\ \end{smallmatrix} &
\begin{smallmatrix} ((\boldsymbol{\phi}_t \odot \tilde{\boldsymbol{\phi}}_t - \boldsymbol{\phi}_t) \otimes \textbf{x}_{A,t})\boldsymbol{\delta}_{t_U,t}^H \\ + ((\boldsymbol{\phi}_t \odot \tilde{\boldsymbol{\phi}}_t) \otimes \boldsymbol{\delta}_{t_A,t}) \boldsymbol{\delta}_{t_U,t}^H \end{smallmatrix} &
\begin{smallmatrix} (\boldsymbol{\phi}_t \odot \tilde{\boldsymbol{\phi}}_t - \boldsymbol{\phi}_t) (\boldsymbol{\phi}_t^H \odot \tilde{\boldsymbol{\phi}}_t^H - \boldsymbol{\phi}_t^H) \otimes \textbf{x}_{A,t} \textbf{x}_{U,t}^H \\ + (\boldsymbol{\phi}_t \odot \tilde{\boldsymbol{\phi}}_t - \boldsymbol{\phi}_t) (\boldsymbol{\phi}_t^H \odot \tilde{\boldsymbol{\phi}}_t^H) \otimes \textbf{x}_{A,t} \boldsymbol{\delta}_{t_U,t}^H \\ + (\boldsymbol{\phi}_t \odot \tilde{\boldsymbol{\phi}}_t) (\boldsymbol{\phi}_t^H \odot \tilde{\boldsymbol{\phi}}_t^H - \boldsymbol{\phi}_t^H) \otimes \boldsymbol{\delta}_{t_A,t} \textbf{x}_{U,t}^H \\ + (\boldsymbol{\phi}_t \odot \tilde{\boldsymbol{\phi}}_t) (\boldsymbol{\phi}_t^H \odot \tilde{\boldsymbol{\phi}}_t^H) \otimes \boldsymbol{\delta}_{t_A,t} \boldsymbol{\delta}_{t_U,t}^H \\ \end{smallmatrix} \\ 
\boldsymbol{\delta}_{t_U,t} \boldsymbol{\delta}_{t_A,t}^H & 
\begin{smallmatrix} \boldsymbol{\delta}_{t_U,t} ((\boldsymbol{\phi}_t^H \odot \tilde{\boldsymbol{\phi}}_t^H - \boldsymbol{\phi}_t^H) \otimes \textbf{x}_{A,t}^H) \\ + \boldsymbol{\delta}_{t_U,t} ((\boldsymbol{\phi}_t^H \odot \tilde{\boldsymbol{\phi}}_t^H) \otimes \boldsymbol{\delta}_{t_A,t}^H) \end{smallmatrix} & 
\boldsymbol{\delta}_{t_U,t} \boldsymbol{\delta}_{t_U,t}^H & 
\begin{smallmatrix} \boldsymbol{\delta}_{t_U,t} ((\boldsymbol{\phi}_t^H \odot \tilde{\boldsymbol{\phi}}_t^H - \boldsymbol{\phi}_t^H) \otimes \textbf{x}_{U,t}^H) \\ + \boldsymbol{\delta}_{t_U,t} ((\boldsymbol{\phi}_t^H \odot \tilde{\boldsymbol{\phi}}_t^H) \otimes \boldsymbol{\delta}_{t_U,t}^H) \end{smallmatrix} \\ 
\begin{smallmatrix} ((\boldsymbol{\phi}_t \odot \tilde{\boldsymbol{\phi}}_t - \boldsymbol{\phi}_t) \otimes \textbf{x}_{U,t})\boldsymbol{\delta}_{t_A,t}^H \\ + ((\boldsymbol{\phi}_t \odot \tilde{\boldsymbol{\phi}}_t) \otimes \boldsymbol{\delta}_{t_U,t}) \boldsymbol{\delta}_{t_A,t}^H \end{smallmatrix} & 
\begin{smallmatrix} (\boldsymbol{\phi}_t \odot \tilde{\boldsymbol{\phi}}_t - \boldsymbol{\phi}_t) (\boldsymbol{\phi}_t^H \odot \tilde{\boldsymbol{\phi}}_t^H - \boldsymbol{\phi}_t^H) \otimes \textbf{x}_{U,t} \textbf{x}_{A,t}^H \\ + (\boldsymbol{\phi}_t \odot \tilde{\boldsymbol{\phi}}_t - \boldsymbol{\phi}_t) (\boldsymbol{\phi}_t^H \odot \tilde{\boldsymbol{\phi}}_t^H) \otimes \textbf{x}_{U,t} \boldsymbol{\delta}_{t_A,t}^H \\ + (\boldsymbol{\phi}_t \odot \tilde{\boldsymbol{\phi}}_t) (\boldsymbol{\phi}_t^H \odot \tilde{\boldsymbol{\phi}}_t^H - \boldsymbol{\phi}_t^H) \otimes \boldsymbol{\delta}_{t_U,t} \textbf{x}_{A,t}^H \\ + (\boldsymbol{\phi}_t \odot \tilde{\boldsymbol{\phi}}_t) (\boldsymbol{\phi}_t^H \odot \tilde{\boldsymbol{\phi}}_t^H) \otimes \boldsymbol{\delta}_{t_U,t} \boldsymbol{\delta}_{t_A,t}^H \\ \end{smallmatrix} & 
\begin{smallmatrix} ((\boldsymbol{\phi}_t \odot \tilde{\boldsymbol{\phi}}_t - \boldsymbol{\phi}_t) \otimes \textbf{x}_{U,t})\boldsymbol{\delta}_{t_U,t}^H \\ + ((\boldsymbol{\phi}_t \odot \tilde{\boldsymbol{\phi}}_t) \otimes \boldsymbol{\delta}_{t_U,t}) \boldsymbol{\delta}_{t_U,t}^H \end{smallmatrix} & 
\begin{smallmatrix} (\boldsymbol{\phi}_t \odot \tilde{\boldsymbol{\phi}}_t - \boldsymbol{\phi}_t) (\boldsymbol{\phi}_t^H \odot \tilde{\boldsymbol{\phi}}_t^H - \boldsymbol{\phi}_t^H) \otimes \textbf{x}_{U,t} \textbf{x}_{U,t}^H \\ + (\boldsymbol{\phi}_t \odot \tilde{\boldsymbol{\phi}}_t - \boldsymbol{\phi}_t) (\boldsymbol{\phi}_t^H \odot \tilde{\boldsymbol{\phi}}_t^H) \otimes \textbf{x}_{U,t} \boldsymbol{\delta}_{t_U,t}^H \\ + (\boldsymbol{\phi}_t \odot \tilde{\boldsymbol{\phi}}_t) (\boldsymbol{\phi}_t^H \odot \tilde{\boldsymbol{\phi}}_t^H - \boldsymbol{\phi}_t^H) \otimes \boldsymbol{\delta}_{t_U,t} \textbf{x}_{U,t}^H \\ + (\boldsymbol{\phi}_t \odot \tilde{\boldsymbol{\phi}}_t) (\boldsymbol{\phi}_t^H \odot \tilde{\boldsymbol{\phi}}_t^H) \otimes \boldsymbol{\delta}_{t_U,t} \boldsymbol{\delta}_{t_U,t}^H \\ \end{smallmatrix} \\
\end{tabu}
\right]$$
\end{table}}
where all random variables are independent. 
Due to $\mathbb{E}[\boldsymbol{\delta}_{t_A,t}]$ and $\mathbb{E}[\boldsymbol{\delta}_{t_U,t}]$ both having zero mean, $\mathbb{E}[\textbf{e}_t\textbf{e}_t^H]$ becomes:

{\tabulinesep=1.2mm
\begin{table}[ht]
\centering
$$= \left[
\begin{tabu}{cccc}
\mathbb{E}[\boldsymbol{\delta}_{t_A,t} \boldsymbol{\delta}_{t_A,t}^H] & 
\mathbb{E}[\boldsymbol{\phi}_t^H \odot \tilde{\boldsymbol{\phi}}_t^H] \otimes \mathbb{E}[\boldsymbol{\delta}_{t_A,t}\boldsymbol{\delta}_{t_A,t}^H] & 
\textbf{0}_{M \times K} & 
\textbf{0}_{M \times KN} \\
\mathbb{E}[\boldsymbol{\phi}_t \odot \tilde{\boldsymbol{\phi}}_t] \otimes \mathbb{E}[\boldsymbol{\delta}_{t_A,t} \boldsymbol{\delta}_{t_A,t}^H] &
\begin{smallmatrix} \mathbb{E}[(\boldsymbol{\phi}_t \odot \tilde{\boldsymbol{\phi}}_t - \boldsymbol{\phi}_t) (\boldsymbol{\phi}_t^H \odot \tilde{\boldsymbol{\phi}}_t^H - \boldsymbol{\phi}_t^H)] \otimes \textbf{x}_{A,t} \textbf{x}_{A,t}^H \\ + \mathbb{E}[(\boldsymbol{\phi}_t \odot \tilde{\boldsymbol{\phi}}_t) (\boldsymbol{\phi}_t^H \odot \tilde{\boldsymbol{\phi}}_t^H)] \otimes \mathbb{E}[\boldsymbol{\delta}_{t_A,t} \boldsymbol{\delta}_{t_A,t}^H] \\ \end{smallmatrix} &
\textbf{0}_{MN \times K} &
\begin{smallmatrix}
\mathbb{E}[(\boldsymbol{\phi}_t \odot \tilde{\boldsymbol{\phi}}_t - \boldsymbol{\phi}_t) (\boldsymbol{\phi}_t^H \odot \tilde{\boldsymbol{\phi}}_t^H - \boldsymbol{\phi}_t^H)] \otimes \textbf{x}_{A,t} \textbf{x}_{U,t}^H \end{smallmatrix} \\
\textbf{0}_{K \times M} & 
\textbf{0}_{K \times MN} & 
\mathbb{E}[\boldsymbol{\delta}_{t_U,t} \boldsymbol{\delta}_{t_U,t}^H] & 
\mathbb{E}[\boldsymbol{\phi}_t^H \odot \tilde{\boldsymbol{\phi}}_t^H] \otimes \mathbb{E}[\boldsymbol{\delta}_{t_U,t} \boldsymbol{\delta}_{t_U,t}^H] \\
\textbf{0}_{KN \times M} & 
\begin{smallmatrix} \mathbb{E}[(\boldsymbol{\phi}_t \odot \tilde{\boldsymbol{\phi}}_t - \boldsymbol{\phi}_t) (\boldsymbol{\phi}_t^H \odot \tilde{\boldsymbol{\phi}}_t^H - \boldsymbol{\phi}_t^H)] \otimes \textbf{x}_{U,t} \textbf{x}_{A,t}^H \end{smallmatrix} & 
\mathbb{E}[\boldsymbol{\phi}_t \odot \tilde{\boldsymbol{\phi}}_t] \otimes \mathbb{E}[\boldsymbol{\delta}_{t_U,t} \boldsymbol{\delta}_{t_U,t}^H] & 
\begin{smallmatrix} \mathbb{E}[(\boldsymbol{\phi}_t \odot \tilde{\boldsymbol{\phi}}_t - \boldsymbol{\phi}_t) (\boldsymbol{\phi}_t^H \odot \tilde{\boldsymbol{\phi}}_t^H - \boldsymbol{\phi}_t^H)] \otimes \textbf{x}_{U,t} \textbf{x}_{U,t}^H \\ + \mathbb{E}[(\boldsymbol{\phi}_t \odot \tilde{\boldsymbol{\phi}}_t) (\boldsymbol{\phi}_t^H \odot \tilde{\boldsymbol{\phi}}_t^H)] \otimes \mathbb{E}[\boldsymbol{\delta}_{t_U,t} \boldsymbol{\delta}_{t_U,t}^H] \\ \end{smallmatrix} \\
\end{tabu}
\right]$$
\end{table}}

We will first calculate the following terms that commonly appear in $\mathbb{E}[\textbf{e}_t\textbf{e}_t^H]$ are:
\begin{align*}
\textbf{M}_1 &=
\mathbb{E}[(\boldsymbol{\phi}_t \odot \tilde{\boldsymbol{\phi}}_t - \boldsymbol{\phi}_t) (\boldsymbol{\phi}_t^H \odot \tilde{\boldsymbol{\phi}}_t^H - \boldsymbol{\phi}_t^H)] \in \mathbb{C}^{N \times N} \\
\textbf{M}_2 &=
\mathbb{E}[(\boldsymbol{\phi}_t \odot \tilde{\boldsymbol{\phi}}_t) (\boldsymbol{\phi}_t^H \odot \tilde{\boldsymbol{\phi}}_t^H)] \in \mathbb{C}^{N \times N}
\end{align*}
Dropping $t$ for easier notation, the $m$th row and $n$th column elements of $\textbf{M}_1$ and $\textbf{M}_2$ are:
\begin{align*}
[\textbf{M}_1]_{m,n} &= \mathbb{E}[\phi_{m} \tilde{\phi}_{m} \phi_{n}^* \tilde{\phi}_{n}^* - \phi_{m} \tilde{\phi}_{m} \phi_{n}^* - \phi_{m} \phi_{n}^* \tilde{\phi}_{n}^* + \phi_{m} \phi_{n}^*] \\
&= \phi_{m} \phi_{n}^* \mathbb{E}[\tilde{\phi}_{m} \tilde{\phi}_{n}^* - \tilde{\phi}_{m} - \tilde{\phi}_{n}^* + 1] \\
&= \phi_{m} \phi_{n}^* \mathbb{E}[e^{j(\tilde{\theta}_{m} - \tilde{\theta}_{n})} - e^{j\tilde{\theta}_{m}} - e^{-j\tilde{\theta}_{n}} + 1] \\
[\textbf{M}_2]_{m,n} &= \mathbb{E}[\phi_{m} \tilde{\phi}_{m} \phi_{n}^* \tilde{\phi}_{n}^*] \\
&= \phi_{m} \phi_{n}^* \mathbb{E}[e^{j(\tilde{\theta}_{m} - \tilde{\theta}_{n})}]
\end{align*}
The diagonal terms ($m = n$) are
\begin{align*}
[\textbf{M}_1]_{m,m} &= \phi_{m} \phi_{m}^* \mathbb{E}[e^{j(\tilde{\theta}_{m} - \tilde{\theta}_{m})} - e^{j\tilde{\theta}_{m}} - e^{-j\tilde{\theta}_{m}} + 1] \\
&= ||\phi_{m}||^2 (2 - 2\varphi) \\
[\textbf{M}_2]_{m,m} &= \phi_{m} \phi_{m}^* \mathbb{E}[e^{j(\tilde{\theta}_{m} - \tilde{\theta}_{m})}] \\
&= ||\phi_{m}||^2
\end{align*}
while the off diagonal terms ($m \ne n$) are
\begin{align*}
[\textbf{M}_1]_{m,n} &= \phi_{m} \phi_{n}^* \mathbb{E}[e^{j\tilde{\theta}_{m}}e^{-j\tilde{\theta}_{n}} - e^{j\tilde{\theta}_{m}} - e^{-j\tilde{\theta}_{n}} + 1] \\
&= \phi_{m} \phi_{n}^* (\varphi^2 - 2\varphi + 1) \\
[\textbf{M}_2]_{m,n} &= \phi_{m} \phi_{n}^* \mathbb{E}[e^{j\tilde{\theta}_{m}}e^{-j\tilde{\theta}_{n}}] \\
&= \phi_{m} \phi_{n}^* \varphi^2
\end{align*}
Therefore,
\begin{align*}
\textbf{M}_1 &= (\varphi^2 - 2\varphi + 1) \boldsymbol{\phi}_t \boldsymbol{\phi}_t^H + (1 - \varphi^2) \textbf{I}_N \\
\textbf{M}_2 &= \varphi^2 \boldsymbol{\phi}_t \boldsymbol{\phi}_t^H + (1 - \varphi^2) \textbf{I}_N
\end{align*}

Substituting in the solved terms for $\textbf{M}_1$ and $\textbf{M}_2$, we obtain the correlation matrix:
{\tabulinesep=1.2mm
\begin{table}[ht]
\centering
$$\mathbb{E}[\textbf{e}_t\textbf{e}_t^H] = \left[
\begin{tabu}{cccc}
\boldsymbol{\Sigma}_{t_A} & 
\varphi(\boldsymbol{\phi}_t^H \otimes \boldsymbol{\Sigma}_{t_A}) & 
\textbf{0}_{M \times K} & 
\textbf{0}_{M \times KN} \\ 
\varphi(\boldsymbol{\phi}_t \otimes \boldsymbol{\Sigma}_{t_A}) &
\begin{smallmatrix} ((\varphi^2 - 2\varphi + 1) \boldsymbol{\phi}_t \boldsymbol{\phi}_t^H + (1 - \varphi^2) \textbf{I}_N) \otimes \textbf{x}_{A,t} \textbf{x}_{A,t}^H \\ + (\varphi^2 \boldsymbol{\phi}_t \boldsymbol{\phi}_t^H + (1 - \varphi^2) \textbf{I}_N) \otimes \boldsymbol{\Sigma}_{t_A} \\ \end{smallmatrix} &
\textbf{0}_{MN \times K} &
\begin{smallmatrix}
((\varphi^2 - 2\varphi + 1) \boldsymbol{\phi}_t \boldsymbol{\phi}_t^H + (1 - \varphi^2) \textbf{I}_N) \otimes \textbf{x}_{A,t} \textbf{x}_{U,t}^H \end{smallmatrix} \\ 
\textbf{0}_{K \times M} & 
\textbf{0}_{K \times MN} & 
\boldsymbol{\Sigma}_{t_U} & 
\varphi(\boldsymbol{\phi}_t^H \otimes \boldsymbol{\Sigma}_{t_U}) \\ 
\textbf{0}_{KN \times M} & 
\begin{smallmatrix} ((\varphi^2 - 2\varphi + 1) \boldsymbol{\phi}_t \boldsymbol{\phi}_t^H + (1 - \varphi^2) \textbf{I}_N) \otimes \textbf{x}_{U,t} \textbf{x}_{A,t}^H \end{smallmatrix} & 
\varphi(\boldsymbol{\phi}_t \otimes \boldsymbol{\Sigma}_{t_U}) & 
\begin{smallmatrix} ((\varphi^2 - 2\varphi + 1) \boldsymbol{\phi}_t \boldsymbol{\phi}_t^H + (1 - \varphi^2) \textbf{I}_N) \otimes \textbf{x}_{U,t} \textbf{x}_{U,t}^H \\ + (\varphi^2 \boldsymbol{\phi}_t \boldsymbol{\phi}_t^H + (1 - \varphi^2) \textbf{I}_N) \otimes \boldsymbol{\Sigma}_{t_U} \\ \end{smallmatrix} \\
\end{tabu}
\right]$$
\end{table}}

% end Appendix A

\section{Diagonal Matrix with pilots and RIS phase values}
To show why the choice of pilots and RIS phase values will make $(\textbf{X}^H\textbf{X} +
\textbf{X}^H \mathbb{E}[\textbf{E}] +
\mathbb{E}[\textbf{E}^H] \textbf{X}  + 
\mathbb{E}[\textbf{E}^H\textbf{E}])$ a diagonal matrix, we can expand out each term into the described pilots and RIS phase shifts.
All terms that result in zero due to the summation over $T$ transmissions are indicated with \textcolor{blue}{blue} colouring and all terms that result in being a diagonal matrix are indicated with \textcolor{red}{red} colouring.

{\fontsize{8}{8} \selectfont
\begin{flalign}
    & \textbf{X}^H\textbf{X} =
    \sum_{t=1}^T (\textbf{x}_t \textbf{x}_t^H)^*
    \otimes \textcolor{red}{\textbf{I}_M} \\
    &= \sum_{t=1}^T 
    \begin{bmatrix}
        \textcolor{red}{\textbf{x}_{A,t} \textbf{x}_{A,t}^H} & \textcolor{blue}{\boldsymbol{\phi}_t^H} \otimes \textcolor{red}{\textbf{x}_{A,t} \textbf{x}_{A,t}^H} & \textcolor{blue}{\textbf{x}_{A,t} \textbf{x}_{U,t}^H} & \textcolor{blue}{\boldsymbol{\phi}_t^H} \otimes \textcolor{blue}{\textbf{x}_{A,t} \textbf{x}_{U,t}^H} \\
        \textcolor{blue}{\boldsymbol{\phi}_t} \otimes \textcolor{red}{\textbf{x}_{A,t} \textbf{x}_{A,t}^H} & \textcolor{red}{\boldsymbol{\phi}_t \boldsymbol{\phi}_t^H} \otimes \textcolor{red}{\textbf{x}_{A,t} \textbf{x}_{A,t}^H} & \textcolor{blue}{\boldsymbol{\phi}_t} \otimes \textcolor{blue}{\textbf{x}_{A,t} \textbf{x}_{U,t}^H}  & \textcolor{red}{\boldsymbol{\phi}_t \boldsymbol{\phi}_t^H} \otimes \textcolor{blue}{\textbf{x}_{A,t} \textbf{x}_{U,t}^H} \\
        \textcolor{blue}{\textbf{x}_{U,t} \textbf{x}_{A,t}^H} & \textcolor{blue}{\boldsymbol{\phi}_t^H} \otimes \textcolor{blue}{\textbf{x}_{U,t} \textbf{x}_{A,t}^H} & \textcolor{red}{\textbf{x}_{U,t} \textbf{x}_{U,t}^H} & \textcolor{blue}{\boldsymbol{\phi}_t^H} \otimes \textcolor{red}{\textbf{x}_{U,t} \textbf{x}_{U,t}^H} \\
        \textcolor{blue}{\boldsymbol{\phi}_t} \otimes \textcolor{blue}{\textbf{x}_{U,t} \textbf{x}_{A,t}^H} & \textcolor{red}{\boldsymbol{\phi}_t \boldsymbol{\phi}_t^H} \otimes \textcolor{blue}{\textbf{x}_{U,t} \textbf{x}_{A,t}^H} & \textcolor{blue}{\boldsymbol{\phi}_t} \otimes \textcolor{red}{\textbf{x}_{U,t} \textbf{x}_{U,t}^H} & \textcolor{red}{\boldsymbol{\phi}_t \boldsymbol{\phi}_t^H} \otimes \textcolor{red}{\textbf{x}_{U,t} \textbf{x}_{U,t}^H}
    \end{bmatrix}^*
    \otimes \textcolor{red}{\textbf{I}_M} \notag
\end{flalign}
\begin{flalign}
    &\textbf{X}^H \mathbb{E}[\textbf{E}] =
    \sum_{t=1}^T (\textbf{x}_t \mathbb{E}[\textbf{e}_t^H])^*
    \otimes \textcolor{red}{\textbf{I}_M} \\
    &= \sum_{t=1}^T 
    \begin{bmatrix}
        \textbf{x}_{A,t} \textcolor{blue}{\textbf{0}_{1 \times M}} & ((\varphi - 1) \textcolor{blue}{\boldsymbol{\phi}_t^H} ) \otimes \textcolor{red}{\textbf{x}_{A,t} \textbf{x}_{A,t}^H} & \textbf{x}_{A,t} \textcolor{blue}{\textbf{0}_{1 \times K}}  & ((\varphi - 1) \textcolor{blue}{\boldsymbol{\phi}_t^H} ) \otimes \textcolor{blue}{\textbf{x}_{A,t} \textbf{x}_{U,t}^H} \\
        \textcolor{blue}{\boldsymbol{\phi}_t} \otimes \textbf{x}_{A,t} \textcolor{blue}{\textbf{0}_{1 \times M}} &  ((\varphi - 1) \textcolor{red}{\boldsymbol{\phi}_t \boldsymbol{\phi}_t^H} ) \otimes \textcolor{red}{\textbf{x}_{A,t} \textbf{x}_{A,t}^H} & \textcolor{blue}{\boldsymbol{\phi}_t} \otimes \textbf{x}_{A,t} \textcolor{blue}{\textbf{0}_{1 \times K}} & ((\varphi - 1) \textcolor{red}{\boldsymbol{\phi}_t \boldsymbol{\phi}_t^H} ) \otimes \textcolor{blue}{\textbf{x}_{A,t} \textbf{x}_{U,t}^H} \\
        \textbf{x}_{U,t} \textcolor{blue}{\textbf{0}_{1 \times M}} & ((\varphi - 1) \textcolor{blue}{\boldsymbol{\phi}_t^H} ) \otimes \textcolor{blue}{\textbf{x}_{U,t} \textbf{x}_{A,t}^H} & \textbf{x}_{U,t} \textcolor{blue}{\textbf{0}_{1 \times K}} & ((\varphi - 1) \textcolor{blue}{\boldsymbol{\phi}_t^H} ) \otimes \textcolor{red}{\textbf{x}_{U,t} \textbf{x}_{U,t}^H} \\
        \textcolor{blue}{\boldsymbol{\phi}_t} \otimes \textbf{x}_{U,t} \textcolor{blue}{\textbf{0}_{1 \times M}} & ((\varphi - 1) \textcolor{red}{\boldsymbol{\phi}_t \boldsymbol{\phi}_t^H}) \otimes \textcolor{blue}{\textbf{x}_{U,t} \textbf{x}_{A,t}^H} & \textcolor{blue}{\boldsymbol{\phi}_t} \otimes \textbf{x}_{U,t} \textcolor{blue}{\textbf{0}_{1 \times K}} & ((\varphi - 1) \textcolor{red}{\boldsymbol{\phi}_t \boldsymbol{\phi}_t^H} ) \otimes \textcolor{red}{\textbf{x}_{U,t} \textbf{x}_{U,t}^H} 
    \end{bmatrix}^*
    \otimes \textcolor{red}{\textbf{I}_M} \notag
\end{flalign}
\begin{flalign}
    & \mathbb{E}[\textbf{E}^H]\textbf{X} =
    \sum_{t=1}^T (\mathbb{E}[\textbf{e}_t] \textbf{x}_t^H)^*
    \otimes \textcolor{red}{\textbf{I}_M} \\
    &= \sum_{t=1}^T 
    \begin{bmatrix}
        \textcolor{blue}{\textbf{0}_{M \times 1}} \textbf{x}_{A,t}^H & \textcolor{blue}{\boldsymbol{\phi}_t^H} \otimes \textcolor{blue}{\textbf{0}_{M \times 1}} \textbf{x}_{A,t}^H & \textcolor{blue}{\textbf{0}_{M \times 1}} \textbf{x}_{U,t}^H & \textcolor{blue}{\boldsymbol{\phi}_t^H} \otimes \textcolor{blue}{\textbf{0}_{M \times 1}} \textbf{x}_{U,t}^H \\
        ((\varphi - 1) \textcolor{blue}{\boldsymbol{\phi}_t}) \otimes \textcolor{red}{\textbf{x}_{A,t} \textbf{x}_{A,t}^H} & ((\varphi - 1) \textcolor{red}{\boldsymbol{\phi}_t \boldsymbol{\phi}_t^H}) \otimes \textcolor{red}{\textbf{x}_{A,t} \textbf{x}_{A,t}^H} & ((\varphi - 1) \textcolor{blue}{\boldsymbol{\phi}_t}) \otimes \textcolor{blue}{\textbf{x}_{A,t} \textbf{x}_{U,t}^H}  & ((\varphi - 1) \textcolor{red}{\boldsymbol{\phi}_t \boldsymbol{\phi}_t^H}) \otimes \textcolor{blue}{\textbf{x}_{A,t} \textbf{x}_{U,t}^H} \\
        \textcolor{blue}{\textbf{0}_{K \times 1}} \textbf{x}_{A,t}^H & \textcolor{blue}{\boldsymbol{\phi}_t^H} \otimes \textcolor{blue}{\textbf{0}_{K \times 1}} \textbf{x}_{A,t}^H & \textcolor{blue}{\textbf{0}_{K \times 1}} \textbf{x}_{U,t}^H & \textcolor{blue}{\boldsymbol{\phi}_t^H} \otimes \textcolor{blue}{\textbf{0}_{K \times 1}} \textbf{x}_{U,t}^H \\
        ((\varphi - 1) \textcolor{blue}{\boldsymbol{\phi}_t}) \otimes \textcolor{blue}{\textbf{x}_{U,t} \textbf{x}_{A,t}^H} & ((\varphi - 1) \textcolor{red}{\boldsymbol{\phi}_t \boldsymbol{\phi}_t^H}) \otimes \textcolor{blue}{\textbf{x}_{U,t} \textbf{x}_{A,t}^H} & ((\varphi - 1) \textcolor{blue}{\boldsymbol{\phi}_t}) \otimes \textcolor{red}{\textbf{x}_{U,t} \textbf{x}_{U,t}^H} & ((\varphi - 1) \textcolor{red}{\boldsymbol{\phi}_t \boldsymbol{\phi}_t^H}) \otimes \textcolor{red}{\textbf{x}_{U,t} \textbf{x}_{U,t}^H} 
    \end{bmatrix}^*
    \otimes \textcolor{red}{\textbf{I}_M} \notag
\end{flalign}
\begin{flalign}
    & \mathbb{E}[\textbf{E}^H\textbf{E}] =
    \sum_{t=1}^T (\mathbb{E}[\textbf{e}_t \textbf{e}_t^H])^*
    \otimes \textcolor{red}{\textbf{I}_M} \label{eq:E_EEH_expanded}\\
    &= \sum_{t=1}^T 
    \begin{bmatrix}
        \textcolor{red}{\boldsymbol{\Sigma}_{t_A}} & \varphi \textcolor{blue}{\boldsymbol{\phi}_t^H} \otimes \textcolor{red}{\boldsymbol{\Sigma}_{t_A}} & \textcolor{blue}{\textbf{0}_{M \times K}} & \textcolor{blue}{\textbf{0}_{M \times KN}} \\
        \varphi \textcolor{blue}{\boldsymbol{\phi}_t} \otimes \textcolor{red}{\boldsymbol{\Sigma}_{t_A}} & \textcolor{red}{\textbf{A}_t} & \textcolor{blue}{\textbf{0}_{MN \times K}} & \textcolor{blue}{\textbf{B}_t} \\
        \textcolor{blue}{\textbf{0}_{K \times M}} & \textcolor{blue}{\textbf{0}_{K \times MN}} & \textcolor{red}{\boldsymbol{\Sigma}_{t_U}} & \varphi \textcolor{blue}{\boldsymbol{\phi}_t^H} \otimes \textcolor{red}{\boldsymbol{\Sigma}_{t_U}} \\
        \textcolor{blue}{\textbf{0}_{KN \times M}} & \textcolor{blue}{\textbf{C}_t} & \varphi \textcolor{blue}{\boldsymbol{\phi}_t} \otimes \textcolor{red}{\boldsymbol{\Sigma}_{t_U}} & \textcolor{red}{\textbf{D}_t}
    \end{bmatrix}^*
    \otimes \textcolor{red}{\textbf{I}_M} \notag
\end{flalign}
}

where
\begin{align}
    \textcolor{red}{\textbf{A}_t} &= ((\varphi^2 - 2\varphi + 1) \textcolor{red}{\boldsymbol{\phi}_t\boldsymbol{\phi}_t^H} + (1 - \varphi^2) \textcolor{red}{\textbf{I}_N}) \otimes \textcolor{red}{\textbf{x}_{A,t} \textbf{x}_{A,t}^H} + (\varphi^2 \textcolor{red}{\boldsymbol{\phi}_t\boldsymbol{\phi}_t^H} + (1-\varphi^2)\textcolor{red}{\textbf{I}_N}) \otimes \textcolor{red}{\boldsymbol{\Sigma}_{t_A}} \notag \\
    \textcolor{blue}{\textbf{B}_t} & = ((\varphi^2 - 2\varphi + 1) \textcolor{red}{\boldsymbol{\phi}_t\boldsymbol{\phi}_t^H} + (1 - \varphi^2) \textcolor{red}{\textbf{I}_N}) \otimes \textcolor{blue}{\textbf{x}_{A,t} \textbf{x}_{U,t}^H} \notag \\
    \textcolor{blue}{\textbf{C}_t} & = ((\varphi^2 - 2\varphi + 1) \textcolor{red}{\boldsymbol{\phi}_t\boldsymbol{\phi}_t^H} + (1 - \varphi^2) \textcolor{red}{\textbf{I}_N}) \otimes \textcolor{blue}{\textbf{x}_{U,t} \textbf{x}_{A,t}^H} \notag \\
    \textcolor{red}{\textbf{D}_t} & = ((\varphi^2 - 2\varphi + 1) \textcolor{red}{\boldsymbol{\phi}_t\boldsymbol{\phi}_t^H} + (1 - \varphi^2) \textcolor{red}{\textbf{I}_N}) \otimes \textcolor{red}{\textbf{x}_{U,t} \textbf{x}_{U,t}^H} + (\varphi^2 \textcolor{red}{\boldsymbol{\phi}_t\boldsymbol{\phi}_t^H} + (1-\varphi^2)\textcolor{red}{\textbf{I}_N}) \otimes \textcolor{red}{\boldsymbol{\Sigma}_{t_U}} \notag 
\end{align}

} % end Appendices

\end{document}